\documentclass[conference]{IEEEtran}
\usepackage{cite}
\usepackage{amsmath}
\usepackage{enumitem}
\usepackage{amssymb}
\usepackage{amsthm}
\usepackage{graphicx,epstopdf}
\usepackage{multicol}
\usepackage{textcomp}
\usepackage{multirow}
\usepackage{flushend}
\newcommand\T{\rule{0pt}{2.4ex}}

\usepackage{color}
\makeatletter
\newcommand*\bigcdot{\mathpalette\bigcdot@{0.85}}
\newcommand*\bigcdot@[2]{\mathbin{\vcenter{\hbox{\scalebox{#2}{$\m@th#1\bullet$}}}}}
\makeatother

\begin{document}
\title{Sizing of Movable Energy Resources for Service Restoration and Reliability Enhancement\\}

\author{Narayan Bhusal, \emph{Student Member, IEEE}, Mukesh Gautam, \emph{Student Member, IEEE}, \\ and Mohammed Benidris, \emph{Member, IEEE}\\
Department of Electrical and Biomedical Engineering\\ 
University of Nevada, Reno, NV 89557, USA\\
Emails: bhusalnarayan62@nevada.unr.edu, mukesh.gautam@nevada.unr.edu, and  mbenidris@unr.edu} 

\maketitle

\begin{abstract}
The frequency of extreme events (e.g., hurricanes, earthquakes, and floods) and man-made attacks (cyber and physical attacks) has increased dramatically in recent years. These events have severely impacted power systems ranging from long outage times to major equipment (e.g., substations, transmission lines, power plants, and distribution system) destruction. Distribution system failures and outages are major contributors to power supply interruptions. Network reconfiguration and movable energy resources (MERs) can play a vital role in supplying loads during and after contingencies. This paper proposes a two-stage strategy to determine the minimum sizes of MERs with network reconfiguration for distribution service restoration and supplying local and isolated loads. Sequential Monte Carlo simulations are used to model the outages of distribution system components. After a contingency, the first stage determines the network reconfiguration based on the spanning tree search algorithm. In the second stage, if some system loads cannot be fed by network reconfiguration, MERs are deployed and the optimal routes to reach isolated areas are determined based on the Dijkstra’s shortest path algorithm (DSPA).  The traveling time obtained from the DSPA is incorporated with the proposed sequential Monte Carlo simulation-based approach to determine the sizes of MERs. The proposed method is applied on several distribution systems including the IEEE-13 and IEEE-123 node test feeders. The results show that network reconfiguration can reduce the required sizes of MERs to supply the isolated areas.
\end{abstract}
\IEEEpeerreviewmaketitle

\section{INTRODUCTION}
The frequency of extreme weather events (e.g., hurricanes, earthquakes, and floods) has been increasing over the past a few years. For example, the average number of disaster events in the United States from 2014 to 2018 is more than double the average number of disaster events from 1980 to 2018 \cite{NONAA19}. These events have led to large blackouts and major destructions of power grids resulting in economic losses and more importantly, long outage duration times. The super-storm Sandy of October 2012 caused power loss to 8 million customers across 15 states in the United States \cite{USDOE1328}. Hurricane Irene in 2011 caused power outage for 6.5 million people \cite{USDOE1328}. Hurricane Harvey in 2017 caused power outage to more than 2 million customers \cite{NERC1846}. Majority of the customer interruptions are caused by the distribution system failures\cite{RBill1996, RBrown2002, AChowdhury2009}. Therefore, this calls for developing control and operation methods and planning strategies that can improve distribution service restoration (DSR) after such events.  

Typically, DSR is performed after contingencies to maximize load restoration and minimize outage durations. Although numerous approaches have been documented in the literature for DSR such as network reconfiguration (NR), microgrid (MG) formation, and splitting power grid into several smaller and reliable MG, coordination between NR, Distributed Generations (DGs), and movable energy resources (MERs) to form dynamic MGs have been considered as the most effective DSR approach. 

Mobile energy resources (MERs) are flexible and movable resources that can play an important role in distribution service restoration, specifically when there are no other means of power supply during contingencies. MERs can be easily and quickly integrated into distribution systems when sustained damages lead to prolonged power outages. These resources are carried from staging locations to faulted locations with the help of trucks. MERs are not only flexible in terms of carrying them from one location to another but also they can be made sufficiently large to supply large loads in the range of MWs. The feasibility of MERs for distribution service restoration has been extensively studied in the literature. Authors of  \cite{8731929} have proposed a strategy to restore critical loads considering dispatch of repair crew and MERs. An optimization method to form dynamic MG through re-routing of MERs has been introduced in \cite{8476200} to reduce the amount of load shedding caused by extreme events. In \cite{8768215}, mobile energy storage systems (MESSs) have been integrated with MG resources, and distribution network reconfiguration has been performed to minimize the total system cost. Also, damage and repair status of distribution and transportation networks have been considered in the integrated approach of \cite{8768215}. In \cite{7559799}, mobile emergency generators have been dispatched (pre-positioning and real-time) to supply critical loads by forming MG in distribution system. In \cite{8642442}, repair crew, mobile power sources, and other distribution service restoration strategies are coordinated to restore loads after extreme events. Transportable energy storage, generation rescheduling, and network reconfiguration are integrated to enhance the resilience of distribution systems in \cite{8334269}. Most of the current work in the literature is focused on strategies to deploy MERs without determining the required sizes for a spectrum of potential contingencies. Without appropriate size of MERs, it will either be oversized or undersized and therefore may not be able to achieve its objective of maximum service restoration with minimum resources during contingencies.

This paper proposes a stochastic strategy to determine sizes of MERs for distribution service restoration considering network reconfiguration, travel times, and outage durations of distribution system components (switches, transformers, and lines). In the proposed approach, after the occurrence of each contingency, the spanning tree search algorithm (STSA) is applied as a decision tool to check the possibility of network reconfiguration. Based on the statuses of the network configuration, the necessity of MERs is decided and optimal routes for the relocation of MERs are determined based on Dijkstra's shortest path algorithm (DSPA). Considering the traveling time for shortest routes provided by the DSPA and the installation time of MERs, the sequential Monte Carlo simulation-based approach is used to determine the sizes of MERs to serve local and isolated loads. Traveling and installation times are subtracted from contingency durations in determining the sizes of MERs. The effectiveness of the proposed approach is demonstrated through several case studies on IEEE 13- and IEEE 123-node test feeders. The results show that network reconfiguration can reduce the required sizes of MERs to supply isolated loads.

The rest of the paper is organized as follows. Section \ref{reconfiguration} describes the network reconfiguration problem and STSA used for network reconfiguration. Section \ref{sizing} describes the proposed approach to determine the optimal routes for the relocation of MERs, unbalanced power flow, and Monte Carlo simulation-based approach for sizing of MERs. Section \ref{CaseStudies} illustrates the proposed method on the IEEE 13- and IEEE 123-node test feeders. Finally, section \ref{Conclusion} provides concluding remarks.
 
\section{Distribution System Network Reconfiguration}\label{reconfiguration}
Electric distribution systems are characterized by radial or weakly-meshed structure with high $R/X$ branch ratios. Distribution systems are generally equipped with two types of switches:  sectionalizing  switches (normally closed)  and  tie-switches (normally open) to serve the maximum loads during the normal and the contingency conditions. Optimal distribution network reconfiguration is  one of the  important  operational  tasks,  which  are  performed frequently  on  electric distribution  systems to achieve various objectives. Network reconfiguration is the process of altering  the  configuration  of  the  distribution  system by  changing  the  status  of  the  sectionalizing  and  tie-switches to  achieve desired  objectives. Several methods have been presented in the literature to solve the optimal distribution system reconfiguration problem for different objectives. It was first introduced by Merlin and Back in 1975 for minimizing the active power loss of radial distribution systems \cite{10024470517}.  Of  these  objectives,  distribution service restoration \cite{589664, 686986, 6781027},  reliability improvement \cite{6689344, 6960676}, and loss  minimization  and  load  balancing \cite{193906, 8853468} are  of most  concern. 

In this paper, distribution systems are represented as a graph $G$ with $V$ vertices and $E$ nodes and STSA is implemented as a decision tool to search the possibility of network reconfiguration after the contingencies.


In Graph Theory, a graph represents a structure of composed nodes (or vertices) and edges. The edges are the connection between vertices. The electric distribution system also consists of nodes and lines similar to that of the graph, therefore, STSA can also be used for distribution systems\cite{6781027}. Moreover, it is easier to maintain the radiality of the distribution system using STSA than other available techniques.

A spanning tree is a subset of the graph, which has a minimum number of edges connecting all nodes. A spanning tree does not have loops and it is not disconnected. A connected graph can have several spanning trees and all possible spanning trees will have same number of edges and nodes. If all the nodes are connected in a radial distribution system, it will obviously represent a spanning tree. STSA is a highly useful method for reconfiguration of distribution networks. In this method, firstly, all the possible combinations of spanning tree are determined. After this, the optimal spanning tree is selected as per the desired objectives.

\section{Sizing of Movable Energy Resources} \label{sizing}
Sizes of MERs depend upon the curtailed loads, duration of contingencies, the travel time to the location of isolated loads, and installation time. This section describes the application of the DSPA for optimal routes for MERs, calculation of the unbalanced power flow, sizing of MERs, Monte Carlo simulations, and the solution algorithm.

\subsection{DSPA  for Optimal Routing of Movable Energy Resources}
As several routes could be available from the staging position of MERs to the faulted node, finding the best route can definitely reduce the outage duration for the isolated loads. Various approaches such as DSPA \cite{Dijkstra1959} and  Floyd-Warshall algorithm \cite{CormenMcGraw2001} have been used in literature to determine the optimal route. As only the route from staging position to the faulted node is of the interest, in this work, the  DSPA  is used to find the optimal route that minimizes the traveling time.

The DSPA routes the MERs in such a way that the distance and therefore, travel time is minimized. The optimal traveling time obtained from this approach is incorporated in determining the sizes of the MERs that can supply the isolated loads. As MERs may take some time to reach the faulted node from the staging position, ignoring it would produce inaccurate sizes of the MERs. Apart from the traveling time, installation time for MERs at the faulted node is also considered in determining the sizes of MERs.

\subsection{Unbalanced Power Flow and Simulation Environment}
MATLAB is used to simulate potential contingencies, determine optimal sizes and routes, and perform network reconfiguration. For each contingency, if some loads cannot be supplied using network reconfiguration, the OpenDSS is used for power flow calculations which are usually unbalanced. The OpenDSS is an open source for distribution system simulation developed by Electric Power Research Institute (EPRI) \cite{EPRIOpenDSS}. The OpenDSS calculates unbalanced power flow using Newton's Method (note that Newton's Method implemented in OpenDSS is different from the Newton-Raphson method). In integrating MATLAB and OpenDSS, distribution system data including reliability data and routes information are provided in MATLAB. MATLAB calls the OpenDSS engine for the unbalanced power flow. OpenDSS provides all the monitored information back to the MATLAB to perform the remaining tasks.

\subsection{Determination of Sizes of Movable Energy Resources}
\begin{figure}
\vspace{-3ex}
\centering
    \includegraphics[scale=0.88]{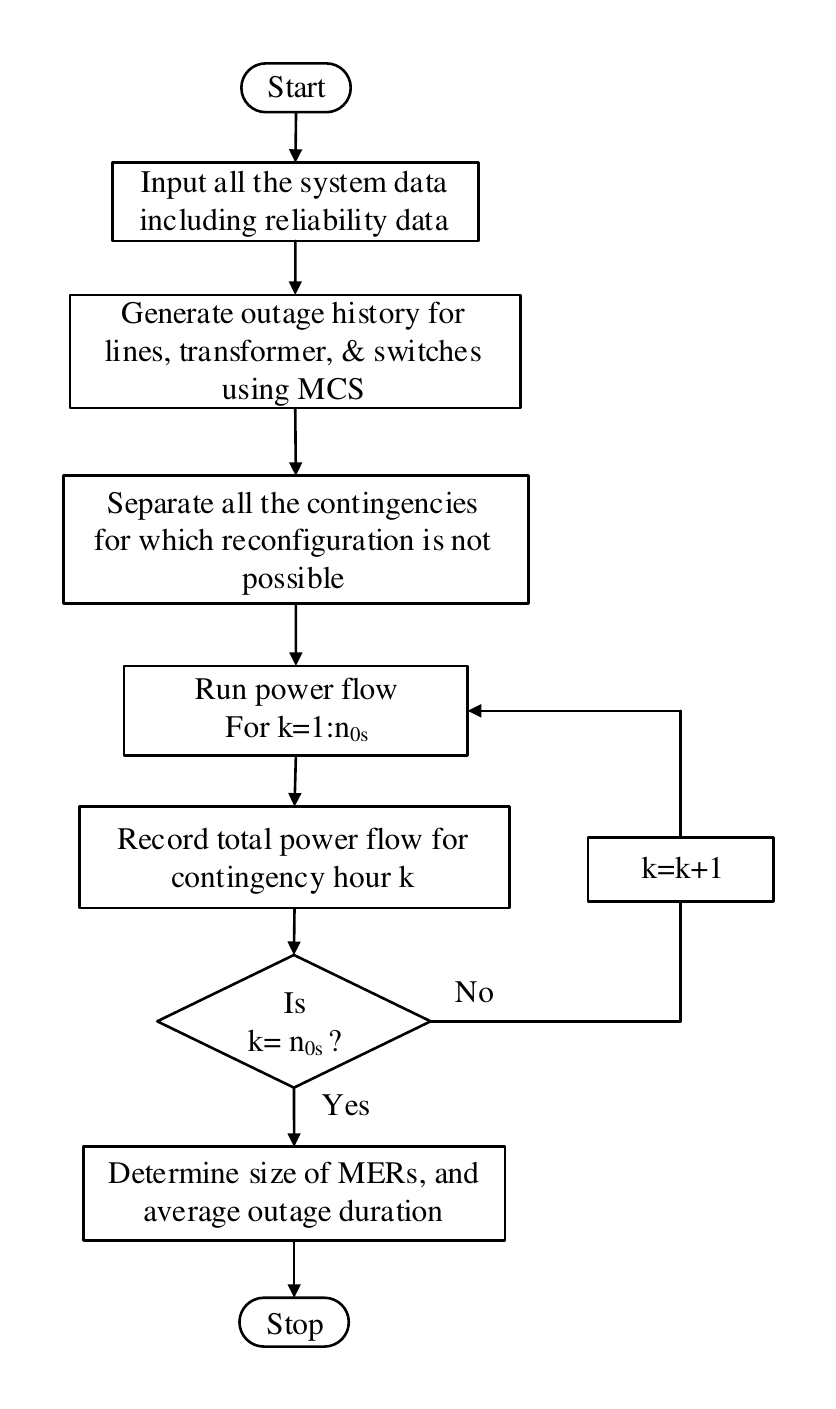}
    \vspace{-3ex}
    \caption{Flow chart of proposed approach}
    \label{fig1}
\end{figure}

When a contingency occurs, several components of the distribution system may become out of service for long periods. In this case, the existing utility supply may not be sufficient or may be out of the reach to certain parts of the distribution system. A portion of these loads could be supplied through reconfiguration and local generations. However, in some cases, even after the reconfiguration and local generations, some areas remain isolated from the main supply. Therefore, no matter how critical these loads are, they will face power outages. To deal with such situations and supply such isolated loads, this work determines the size of MERs that is required to supply those isolated loads. The procedure to determine the sizes of MERs is explained as follows and shown in Fig. \ref{fig1}. 
\begin{enumerate}
    \item Input all the required system data including reliability and routes information.
    \item Perform the power flow for a year for the given system without any line loss and monitor the nodal voltages, currents, and power flows. Calculate the total substation power flow for each hour and store them.
    \item Generate outage history for lines, transformers, and switches using Monte Carlo next event method for a reasonable number of years.
    \item Generate all possible radial configurations using STSA and separate all the contingencies for which network reconfiguration is not possible. 
    \item For each contingency for which spanning tree cannot be formed, perform the power flow with specific hourly loads for the duration of that particular contingency.
    \item Record all substation power flows for the duration of all contingencies. Subtracting this power from the power without any line loss gives the power needed to be supplied by MERs. Note that the time coordination between the total substation power flow due to contingencies and without contingencies with the same level of load is very important otherwise it will give the wrong size.  
    \item Determine the optimal route for the relocation of MERs for each contingency using DSPA. Certain installation time should be added to the optimal travel time to determine the net power requirement. As some time is required for relocation and installation of MERs, the isolated loads will experience outages during this period. This outage duration is subtracted from the duration of a given contingency to determine the average size of the energy storage.
\end{enumerate}

Mathematically, determination of the size of the MERs can be expressed as follows. 
    \begin{equation} \label{eq1}
        P_{net}^k=P_{base}^{k}-P_{after}^k 
    \end{equation}
    \begin{equation}\label{eq2}
        t_{cont}=\sum_{i=1}^{n_{cont}}D_i 
    \end{equation}
    
        \begin{equation} \label{eq3}
           t_{avg}=\frac{t_{cont}}{n_{cont}} 
        \end{equation}
        \begin{equation} \label{eq4}
            E_{net}^i=\sum_{k=1}^{D_i}P_{net}^k
        \end{equation}
        \begin{equation} \label{eq5}
            E_{avg}=\frac{\sum_{i=1}^{t_{cont}} E_{net}^i}{{n_{cont}}} 
        \end{equation}
        \begin{equation} \label{eq6}
            P_{max}=\frac{\sum_{i=1}^{n_{cont}}{P_{max}^i}}{{n_{cont}}} 
        \end{equation}
        \begin{equation} \label{eq7}
           P_{avg}=\frac{\sum_{i=1}^{n_{cont}}{P_{avg}^i}}{{n_{cont}}} 
        \end{equation}

In (\ref{eq1})--(\ref{eq7}), $P_{net}^k$ is the net power required for the MERs after incorporating the traveling and installation time at hour $k$; $P_{base}^k$ is the total substation power flow at hour $k$ without any contingencies; $P_{after}^k$ denotes the total substation power flow after the contingencies and reconfiguration at hour $k$; $t_{cont}$ is the total contingencies duration; $D_i$ is the duration of contingency $i$; $E_{net}^i$ is the sum of power demand during contingency $i$;  $E_{avg}$ is the average size of required MERs;  $n_{cont}$ is the total number of contingencies during $8760\times Year$ hours; $P_{max}$ is the average of the maximum of all the contingencies; $P_{max}^i$ is the maximum power demand from MERs during contingency $i$; $t_{avg}$ is the average duration of the contingencies; $P_{avg}$ denotes the average power needed to supply due to contingencies; and $P_{avg}^i$ is the average power demand from MERs during contingency $i$.

\section{Case Studies}\label{CaseStudies}
In order to validate the proposed approach, simulations have been performed on two distribution feeders: IEEE 13- and IEEE 123-node test feeders. These are very common unbalanced systems for distribution system studies. The load profile is taken from \cite{EPRIOpenDSS}. The reliability data for the tested distribution systems are given in Table \ref{tab:reliability} \cite{6138890}. MATLAB is used for network reconfiguration (spanning-tree), Monte Carlo simulations, and determining the travel time of MERs. The power flow for each scenario is calculated using OpenDSS.
\begin{table}[h!]
\vspace{-1ex}
\caption{Distribution system reliability data\vspace{-1ex}}
    \centering
        \label{tab:reliability}
    \begin{tabular}{c|c|c}
    \hline
        Components & Mean time to failure & Mean time to repair\\
        & (failure/year)& (hour) \T\\
        \hline
         Transformer & $0.05882$  & $144$ \T\\
         Distribution line & $0.13$ & $5$ \T\\
         Switch & $0.2$ & $5$ \T\\
         \hline
    \end{tabular}\vspace{-1ex}
\end{table}

\subsection{IEEE 13-Node Test Feeder}
The IEEE 13-node test feeder is characterized by being small, heavily loaded, unbalanced, contains overhead and underground lines, and has one voltage regulator, shunt capacitors, and an in-line transformer. This feeder operates at $4.16$ kV. The total loads in this system are $3466$ kW and $2102$ kVar. More detailed data of the IEEE 13 node test feeder are given in \cite{IEEEFEEDERS}.

Since this system has only one sectionalizing switch and only one substation, no reconfiguration is possible. Monte Carlo simulations are used to sample occurrence times of contingencies and their durations based on the mean time to failure and mean time to repair of system components. The time at which a contingency occurs and the duration of the contingency are important in determining curtailed loads. For example, if a contingency occurs at $2:00$ pm and lasts for $5$ hours, the MER will be used to feed curtailed loads during this period considering that the system load will be changing during this period. The power flow is solved for each contingency and for each load change during the contingency to determine the curtailed load. For this system, the average interruption time of potential contingencies is found to be $10.8$ hours. The MER sizes and interruption times of the IEEE 13-node test feeder are given in Table \ref{tab:results}.

\subsection{IEEE 123-Node Test Feeder}
\begin{figure}
    \hspace{-4ex}
    \includegraphics[scale=0.8]{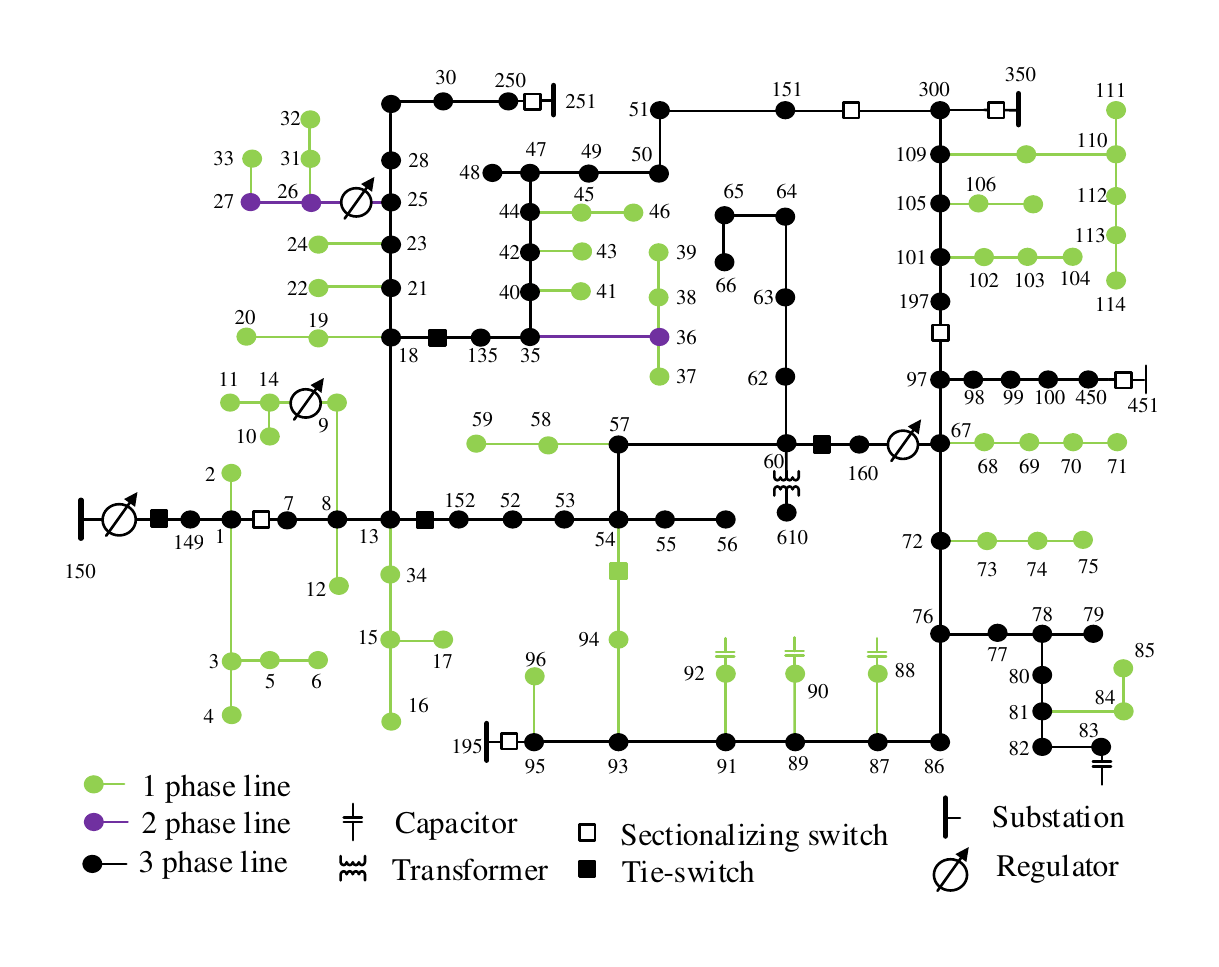}
    \vspace{-5ex}
    \caption{IEEE 123 node-test feeder}
    \vspace{-1ex}
    \label{figieee123}
\end{figure}

The IEEE-123 node test feeder, as shown in Fig. \ref{figieee123}, is characterized by having overhead and underground lines, four voltage regulators, four shunt capacitor banks, multiple sectionalizing and tie-switches, and unbalanced loading with constant current, power, and impedance models. The total real and reactive loads of this system are respectively $3490$ kW and $1925$ kVar. Network data of the IEEE 123-node test feeder are given in \cite{IEEEFEEDERS}.

In this system, there are $6$ sectionalizing switches, $5$ tie-switches, one transformer, and $118$ lines. Reliability data of all network components are used to simulate potential contingencies. During any contingency, the up and down statuses of all system lines, switches, and different substation nodes are provided as inputs to STSA which determines whether the reconfiguration is possible. Power flow calculations are performed only for the contingencies during which load curtailment is not avoidable and reconfiguration is not possible.

In this work, all contingencies for $200$ years are considered. As MER is required only for isolated loads that cannot be fed even by network reconfiguration, power flow calculations will be required only for these cases. Monte Carlo simulations are used to sample occurrence times of contingencies and their durations. For this system, it was found that network reconfiguration alone allows complete load restoration for around $33$\% of the simulated contingencies. For the remaining $67$\% of contingencies, the network reconfiguration can restore some of the loads but some areas remain isolated. Therefore, MERs are required to restore the loads for the cases where the network reconfiguration cannot restore the entire curtailed loads. The average interruption time for this system is found to be $5.84$ hours. The MER sizes and interruption times of the IEEE 13-node test feeder are given in Table \ref{tab:results}.



\subsection{Results and Discussion}
The results of the test cases are shown in Table \ref{tab:results}. The average sizes of MERs for the IEEE 13- and IEEE 123- node test feeders respectively are: ($370$ kW, $3998$ kWh) and ($138$ kW, $810$ kWh). The average of maximum power for each contingency is $590$ kW for IEEE 13-node test feeder and $198$ kW for the IEEE 123- node test feeder. The average durations of each contingency is $10.8$ hours for IEEE 13-node test feeder and $5.84$ hours for the IEEE 123- node test feeder. The travel time and installation time of a MER are subtracted from the interruption time in determining the sizes of MERs since during these times MERs will not deliver energy to isolated loads. Optimal travel time of a MER are calculated using DSPA for the given contingency. The installation time is assumed to be $15$ minutes which represents the time to install MERs at faulted locations. We used $15$ minutes for the sake of simplicity but actual installation times can be used. Even though both the feeders have same level of loads, the average size of MERs in IEEE 13-node is significantly larger than that of the IEEE 123-node because of the fact that reconfiguration is not possible for the case of IEEE 13-node feeder.
\begin{table}[h!]
\vspace{-1ex}
\caption {Results for all case studies\vspace{-1ex}}
\label{tab:results}
\centering
\begin{tabular}{lcc}
\hline
                              &  IEEE 13      &   IEEE 123  \T\\ \hline
Average Size (kWh)            &  $3998$       &   $810$     \T\\
Average Size (kW)             &  $370$        &   $138.67$  \T\\
Maximum Size (kW)             &  $590$        &   $198$     \T\\
Average Duration Time (hours) &  $10.8$       &   $5.84$    \T\\ 
\hline
		\end{tabular}\vspace{-1ex}
\end{table}

\section{Conclusion}\label{Conclusion}
This paper has introduced an optimal sizing of movable energy resources for distribution service restoration considering network reconfiguration, line outage durations, and relocation time of movable resources. The proposed method takes advantage of network reconfiguration for reducing the capacity of energy resources. The spanning tree search algorithm was implemented for network reconfiguration. The time required for the relocation of movable resources and installation was also considered while sizing of those resources. The optimal travel time for relocation of movable energy resources was determined using the Dijkstra's shortest-path algorithm. For the validation of the proposed algorithm, simulations have been performed on the IEEE 13 and IEEE 123 node test feeders, which are highly unbalanced. The test results have illustrated that the proposed algorithm is suitable for sizing of movable energy resources for distribution service restoration and reliability enhancement.
 
\bibliographystyle{IEEEtran}
\bibliography{References.bib}
\end{document}